\documentclass[aps,prl,superscriptaddress,reprint]{revtex4-2}
\usepackage{amssymb}
\usepackage{amsmath}
\usepackage[latin9]{inputenc}
\usepackage{array}
\usepackage{multirow}
\usepackage{color}
\usepackage{esint}
\usepackage{bm}
\usepackage{color}
\usepackage{bbm}
\usepackage{hyperref}
\usepackage{babel}
\usepackage{titlesec}
\usepackage{graphicx,amssymb,color}
\usepackage{epsfig}
\usepackage{epstopdf}
\usepackage{mathtools}
\usepackage{soul}
\usepackage[dvipsnames]{xcolor}
\usepackage{tikz}
\usepackage{tcolorbox}
\usepackage{graphics,amssymb,amsmath,epsfig,color}

\begin{document}
\title{
Thermodynamics of Spin-Imbalanced Fermi Gases with $\mathrm{SU}(N)$ Symmetric Interaction
}

\author{Chengdong He}
\thanks{These authors contributed equally.}
\affiliation{Department of Physics, The Hong Kong University of Science and Technology, Clear Water Bay, Kowloon, Hong Kong, China}

\author{Xin-Yuan Gao}
\thanks{These authors contributed equally.}
\affiliation{Department of Physics, The Chinese University of Hong Kong, Shatin, Hong Kong, China}

\author{Ka Kwan Pak}
\affiliation{Department of Physics, The Hong Kong University of Science and Technology, Clear Water Bay, Kowloon, Hong Kong, China}

\author{Yu-Jun Liu}
\affiliation{Department of Physics, The Hong Kong University of Science and Technology, Clear Water Bay, Kowloon, Hong Kong, China}

\author{Peng Ren}
\affiliation{Department of Physics, The Hong Kong University of Science and Technology, Clear Water Bay, Kowloon, Hong Kong, China}

\author{Mengbo Guo}
\affiliation{Department of Physics, The Hong Kong University of Science and Technology, Clear Water Bay, Kowloon, Hong Kong, China}

\author{Entong Zhao}
\affiliation{Department of Physics, The Hong Kong University of Science and Technology, Clear Water Bay, Kowloon, Hong Kong, China}

\author{Yangqian Yan}
\email{yqyan@cuhk.edu.hk}
\affiliation{Department of Physics, The Chinese University of Hong Kong, Shatin, Hong Kong, China}
\affiliation{
The Chinese University of Hong Kong Shenzhen Research Institute, Shenzhen, China
}%

\author{Gyu-Boong Jo}
\email{gbjo@ust.hk}
\affiliation{Department of Physics, The Hong Kong University of Science and Technology, Clear Water Bay, Kowloon, Hong Kong, China}

\begin{abstract}
Thermodynamics of degenerate Fermi gases has been extensively studied through various aspects such as Pauli blocking effects, collective modes, BCS superfluidity, and more. Despite this, multi-component fermions with imbalanced spin configurations remain largely unexplored, particularly beyond the two-component scenario. In this work, we generalize the thermodynamic study of SU($N$) fermions to spin-imbalanced configurations based on density fluctuations. Theoretically, we provide closed-form expressions of density fluctuation across all temperature ranges for general spin population setups. Experimentally, after calibrating the measurements with deeply degenerate $^{173}$Yb Fermi gases under spin-balanced configurations ($N\leq$~6), we examine the density fluctuations in spin-imbalanced systems. Specifically, we investigate two-species and four-species configurations to validate our theoretical predictions. Our analysis indicates that interaction enhancement effects can be significant even in highly spin-imbalanced systems. Finally, as an application, we use this approach to examine the decoherence process. Our study provides a deeper understanding of the thermodynamic features of spin-imbalanced multi-component Fermi gases and opens new avenues for exploring complex quantum many-body systems.
\end{abstract}

\maketitle

\textit{Introduction.---}Interacting fermions form the foundation of diverse matter types, spanning vast energy and length scales, from materials and ultracold matter~\cite{bloch2008manybody,inguscio2008ultracold} to nuclei and neutron stars~\cite{bailin1984superfluidity,casalbuoni2004inhomogeneous}. While spin populations are typically equal, imbalanced spin systems frequently occur in various physical systems. These systems have provided fundamental insights into exotic quantum phases, ranging from superconductivity under applied magnetic fields~\cite{quay2013spin,linder2015superconducting} to quark superfluidity in the early universe~\cite{alford2008color}. Two-component ultracold fermions with spin imbalance, in particular, enable the study of numerous phenomena, including exotic superfluidity at unitarity~\cite{zwierlein2006fermionic,partridge2006pairing,kinnunen2006strongly,radzihovsky2010imbalanced,gubbels2013imbalanced}, Fermi polarons~\cite{punk2009polarontomolecule,mora2010normal}, and FFLO states in 1D~\cite{hu2007phase,rizzi2008fuldeferrelllarkinovchinnikov,tezuka2008densitymatrix,liao2010spinimbalance}.

While conventional two-component fermionic systems with $\operatorname{SU}(2)$ symmetry constitute the building blocks of most matter, large spins with enhanced $\operatorname{SU}(N)$ symmetry promise novel quantum phenomena and insights. Recent advances in cold atom experiments and theory have realized such $\operatorname{SU}(N)$ many-body systems~\cite{He.2019}.
These include $\operatorname{SU}(N)$ Fermi liquids~\cite{chitov1995renormalizationgroup,yip2014theory,cheng2017mathrm,He.2020,song2020evidence,sonderhouse2020thermodynamics,Zhao.2021}, $\operatorname{SU}(N)$ Mott insulators~\cite{assaraf1999metalinsulator,taie2012su,nataf2016chiral,hofrichter2016direct,zhou2016mott,wang2019slater}, BCS pairing in $N$-component systems~\cite{modawi1997properties,honerkamp2004bcs,he2006superfluidity,rapp2007color,cherng2007superfluidity,ozawa2010population,yip2011theory}, two-orbital $\operatorname{SU}(N)$ fermions~\cite{gorshkov2010twoorbital,zhang2014spectroscopic,scazza2014observation,bois2015phase,bois2016onedimensional,ueda2018symmetry}, and anti-ferromagnetic correlations enhanced by $\operatorname{SU}(N)$ symmetry~\cite{cazalilla2009ultracold,gorshkov2010twoorbital,tamura2019ferromagnetism}. Studying thermodynamics, particularly measuring density fluctuations linked to isothermal compressibility through the fluctuation-dissipation theorem, has been crucial in characterizing $\operatorname{SU}(N)$ Fermi gases~\cite{anderson2015pressure,lee2012compressibility,astrakharchik2007fluctuations}. Experiments with deeply degenerate gases ($T/T_F\ll1$) have shown reduced density fluctuations due to Pauli blocking~\cite{sanner2010suppression,muller2010local,tobias2020thermalization}. In the spin-balanced limit, where each species has an equal population, mean-field theory predicts that interaction effects on thermodynamics are enhanced by a factor of $N-1$~\cite{cazalilla2014ultracold,yip2014theory,cheng2017mathrm,sonderhouse2020thermodynamics}.  However, the thermodynamic characterization of spin-imbalanced multi-component fermions with $\mathrm{SU}(N)$ symmetric interaction has presented significant challenges due to the lack of systematic understanding of interaction enhancement in these systems.

In this work, we present both theoretical and experimental studies of density fluctuations in multi-component Fermi gases with $\mathrm{SU}(N)$ symmetric interaction beyond the spin-balanced limit. We show that Hartree-Fock-type terms in the diagrammatic expansion are most enhanced in the large $N$ limit. We provide closed-form formulas for calculating density fluctuations in Fermi gases with $\operatorname{SU}(N)$ symmetric interactions and spin-imbalanced configurations at finite temperatures. Experimentally, we utilize ${}^{173}\mathrm{Yb}$ atoms to create deeply degenerate Fermi gases with $\operatorname{SU}(N)$ symmetric interactions. The number of species is highly tunable from $N=1$ to $6$, with adjustable proportions in the total population. We first revisit the spin-balanced $\operatorname{SU}(N)$ fermions, calibrating density-fluctuation measurements in the $N=1$ system and demonstrating agreement between our experiments and theory in the balanced $\operatorname{SU}(6)$ system. Next, we examine systems with imbalanced populations, specifically two-species and four-species configurations, validating our theoretical predictions for general setups. Leveraging our analytical expression, we also provide additional theoretical results supporting the fact that interaction enhancement effects can be significant in highly spin-imbalanced systems. Finally, we apply density fluctuation measurements to determine decoherence populations in our systems.

\begin{figure}[tb!]
	\centering
	\includegraphics[width=8.6cm]{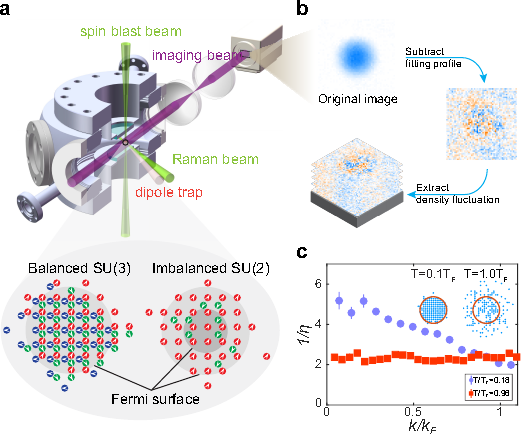}
	\caption{\textbf{Experimental setup for measurement of atom number variance.} (a) Degenerate Fermi gases are prepared in a cross-dipole trap. Momentum distribution of Fermi gas samples is measured using absorption imaging after 20ms TOF time. (b) We subtract a fitting profile for each image to counter total atom number fluctuation. (c) $1/\eta$ for thermal and degenerate Fermi gas at different $k/k_F$. Insets: schematics for state occupation status in the momentum space.}
	\label{atom_variance}
\end{figure}

\begin{figure}[tb!]
	\centering
	\includegraphics[width=8.6cm]{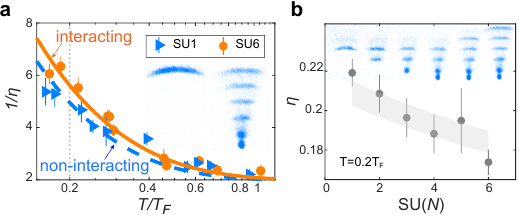}
	\caption{\textbf{Thermometry for interacting degenerate $\mathrm{SU}(N)$ fermions.} 
    (a) Inverse fluctuation $1/\eta$ as a function of the temperature $T$. The orange-solid (blue-dashed) line and orange circles (blue triangles) are the theoretical prediction of $\operatorname{SU}(6)$ [($\operatorname{SU}(1)$] gas multiplied by a factor $\alpha$ and experimental data, respectively. The vertical dotted line denotes the deep-degeneracy temperature $T=0.2T_F$ where data in (b) is measured. 
    (b) Fluctuation $\eta$ for $\operatorname{SU}(N)$ gases as a function of the number of component $N$. The shaded area is the theoretical prediction considering $5\%$ uncertainty in temperature measurement.
    }
	\label{T_dependent}
\end{figure}

\textit{Experimental Procedure.---}Fig.~\ref{atom_variance}(a) shows our experimental setup. We first produce a multi-component degenerate Fermi gas via evaporative cooling in a crossed optical dipole trap (ODT) with two horizontal $1064~\text{nm}$ beams. Using optical pumping and a blast pulse~\cite{song2020evidence}, we prepare atom samples with balanced or imbalanced spin populations. To vary temperatures and densities, we adjust the final depth of the ODT. Finally, we calibrate the trap frequency of all three axes $\{\omega_1,\omega_2,\omega_3\}$ by measuring the dipole mode frequency of the atom cloud.
Spin configurations are calibrated with the optical Stern-Gerlach (OSG) effect; the inset of Fig.~\ref{T_dependent}(b) shows typical OSG images of balanced $\operatorname{SU}(N)$. We then hold the atom cloud in ODT for $400~\text{ms}$ to reach thermal equilibrium, release the atoms, and after $20~\text{ms}$ free expansion, take absorption images with a $3.76$ magnification system.

\textit{Fluctuations Calibration.---}Density fluctuation is measured by counting atom numbers in a small volume from repeated absorption images under the same conditions. We extract the averaged atom number $N_0$ and variance $\Delta N_0^2$, calculating the variance per atom $\eta=\Delta N_0^2/N_0$. To ensure a high signal-to-noise ratio, we use a long image light pulse of about $40~\mu\text{s}$. The imaging system's aberrations and limited resolution (around $4~\mu\text{m}$) cause blurring, making measured $\eta$ lower than the theoretical value~\cite{esteve2006observations,gemelke2009situ}. We bin raw data with specific sizes to reduce these effects. Meanwhile, the bin size cannot be too large due to the limited atom cloud size. We find the variance ratio between degenerate and thermal gases approaches a constant value as the bin size exceeds 7 pixels (See End Note). At that binning size, the extracted variance reflects the actual value of different samples with a constant factor. For all data presented in this work, we use the 7-pixel bin. Prepared atomic samples exhibit fluctuations in atom number and temperature, affecting the extracted $\eta$. We subtract a Fermi fitting profile from each image to exclude these effects and calculate $\eta$ from the subtracted images~\cite{chen2006finitetemperature, butts1997trapped}. A raw and a subtracted image are shown in Fig.~\ref{atom_variance}(b). Red squares and blue circles in Fig.~\ref{atom_variance}(c) show the inverse variance versus momentum for thermal and degenerate Fermi gas, respectively.

Triangles and circles in Fig.~\ref{T_dependent}(a) show the inverse local fluctuation of line-of-sight integrated density profiles measured at the trap center for $\operatorname{SU}(1)$ (i.e., non-interacting system) and $\operatorname{SU}(6)$ \textit{spin-balanced} systems. 
For a non-interacting Fermi gas, it is expected to be $\eta = \alpha\frac{\mathrm{Li}_1(-z_0)}{\mathrm{Li}_2(-z_0)}$, where $\mathrm{Li}$ denotes polylogarithm functions. The fugacity $z_0$ is implicitly given by $-\mathrm{Li}_3(-z_0)=\frac{T_F^3}{6T^3}$, where the Fermi temperature $T_F$ is determined by $T_F = \hbar (\omega_1\omega_2\omega_3)^{\frac{1}{3}}(6N_{0}^{s})^{\frac{1}{3}}$. 
$N_0^{s}=N_0/N$ refers to the number of spins of an arbitrary species. Factor $\alpha$ comes from the blurring effect mentioned above. In our experiment, we extract $\alpha=0.52(2)$ for the best fit (dashed line). For non-interacting $\operatorname{SU}(1)$ gas, as the temperature goes down, the Pauli suppression effect becomes more evident, i.e., the number of available states decreases, and $\eta$ decreases.

To validate the scaling factor $\alpha$, we further compare the theoretical prediction scaled by $\alpha$ with the experimental data of $\operatorname{SU}(6)$ system. Since different spins exert weak repulsive interaction with scattering length $a_sk_F\simeq0.1$ to each other, the expression of $\eta$ of $\operatorname{SU}(1)$ system does not apply anymore. We derive that, for a \textit{homogeneous} $\operatorname{SU}(N)$ Fermi gas, the density fluctuation is approximated by
\begin{equation}
        \eta=\dfrac{-4({T}/{T_F^\text{hom}})^{3/2}\mathrm{Li}_{\frac{1}{2}}\left(-z_0^\text{hom}\right)}{3\sqrt{\pi}-3(N-1){a_sk_F^\text{hom}}\sqrt{{T}/{T_F^\text{hom}}}\mathrm{Li}_{\frac{1}{2}}(-z_0^\text{hom})},
        \label{balanced_fluctuation}
\end{equation}
where $T_F^\text{hom}$, $k_F^\text{hom}$ and $z_0^\text{hom}$ are Fermi temperature, Fermi momentum, and non-interacting fugacity, respectively. 
Equation~(\ref{balanced_fluctuation}) is based on the Hartree-Fock approximation, which becomes exact in the large $N$ limit. 
This is because, in the diagrammatic expansion, we inductively proved that Hartree-Fock terms of the grand potential get $N(N-1)^m$-fold enhancement at order $m$. At the same time, all other diagrams are only enhanced by $N(N-1)^{m'}$ where $m'<m$~\cite{SM}.
It can be checked that in the zero-temperature limit $T/T_F^\text{hom}\rightarrow0$, Eq.~(\ref{balanced_fluctuation}) recovers the well-known relation $\eta=\frac{3}{2}\frac{T/T_F^\text{hom}}{1+\frac{2}{\pi}(N-1)k_F^\text{hom}a_s}$~\cite{yip2014theory,sonderhouse2020thermodynamics,Zhao.2021}.
To apply Eq.~(\ref{balanced_fluctuation}) to our system, we perform local-density approximation and integrate along the line of sight of imaging. Further multiplied by the aforementioned scaling factor $\alpha=0.52$, our theory [red solid line in Fig.~\ref{T_dependent}(a)] agrees well with the experimental data (circles). Compared with $\operatorname{SU}(1)$ gas, the fluctuation for $\operatorname{SU}(6)$ Fermi gas is even lower due to repulsive interaction, and the effect is more pronounced in the deeply degenerate region. Besides temperature, it is clear that the total number of species $N$ also influences the interaction effect. Fig.~\ref{T_dependent}(b) demonstrates how the fluctuation reduces as we gradually increase $N$ from $1$ to $6$. All samples are prepared with approximately the same temperature $T/T_F \simeq 0.2$.
At this temperature, $\eta$ for $\operatorname{SU}(6)$ gas is $16\%$ lower than $\operatorname{SU}(1)$ gas. 

\begin{figure}[tb!]
	\centering
	\includegraphics[width=8.4cm]{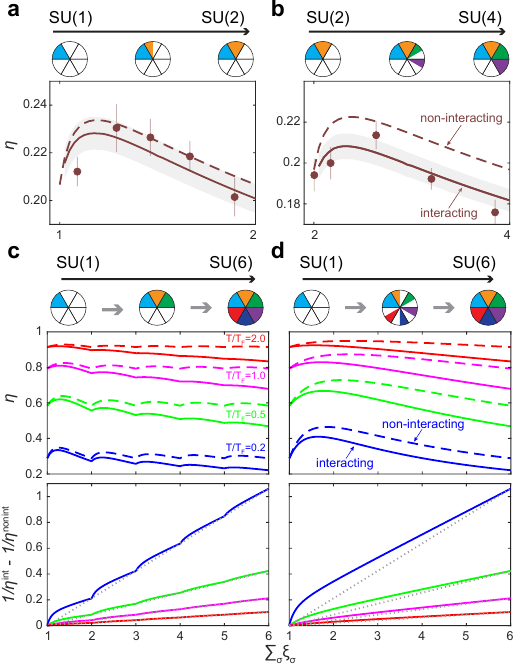}
	\caption{
    \textbf{Density fluctuation $\eta$ of spin-imbalanced systems. }
    (a) Circles show the experimental data of $\eta$ as a function of $\sum_\sigma \xi_\sigma$, with varying $\xi_2$ from 0 to 1. Solid lines and dashed show the theoretical prediction of an interacting and non-interacting system, respectively, with $a_sk_F=0.1$ and $T/T_F=0.2$. The shaded areas denote the error considering $3\%$ temperature uncertainty. 
    (b) Comparing to (a), instead of varying $\xi_2$, $\xi_3$ and $\xi_4$ are simultaneously changed from $0$ to $1$ while $\xi_2$ and $\xi_1$ are kept at 1.
    (c) Theoretical predictions of $\eta$ in homogeneous systems without correcting factor $\alpha$ based on Eq.~(\ref{imbalanced_fluctuation}). The upper panel shows how $\eta$ changes when $\xi_\sigma$ for $\sigma=2,\cdots,6$ are tuned from $0$ to $1$ \textit{one by one} as illustrated by the chart schematic. Solid lines are results of interacting systems $\eta^{\mathrm{int}}$ with $a_sk_F^\mathrm{hom}=0.1$, while dashed lines are $\eta^{\mathrm{nonint}}$ for non-interacting systems with $a_sk_F^\mathrm{hom}=0$. The solid line in the lower panel shows the difference in the inverse fluctuation between interacting and non-interacting systems. The dotted lines show a naive generalization of results from spin-balanced systems. 
    (d) Same plot as (c), except that $\sigma=2,\cdots,6$ are tuned from $0$ to $1$ \textit{simultaneously}.
    }
	\label{Imbalanced}
\end{figure}

\textit{Spin-Imbalanced Gases.---}With fluctuation calibration completed using spin-balanced systems, we now examine spin-imbalanced systems, where numbers of particles in each spin are $N_0^{(\sigma)}=\xi_\sigma N_0^{s}$; $N_0^{s}$ represents the number of particles for the majority spin and $\xi_1=1$. To investigate the influence of an imbalanced population on the system, we study the number fluctuation of $\operatorname{SU}(N)$ system when ``$N$" is smoothly tuned, by which we mean we change $\sum_\sigma \xi_\sigma$ from $N$ to a different value.

We experimentally study the fluctuation in two schemes in the deep quantum degeneracy regime, $T/T_F=0.2$: crossover from $\operatorname{SU}(1)$ to $\operatorname{SU}(2)$ gases by
modifying $\xi_2$ from $0$ to $1$ [circles in Fig.~\ref{Imbalanced}(a)] and crossover from $\operatorname{SU}(2)$ to $\operatorname{SU}(4)$ by simultaneously modifying $\xi_3$ and $\xi_4$ from $0$ to $1$ while keeping $\xi_2=\xi_1=1$ [circles in Fig.~\ref{Imbalanced}(b)].  
We observe a non-monotonic dependence of fluctuation on the number of particles. The quick increase in the highly spin imbalanced regime is because, at a given global equilibrium temperature $T$, the Fermi temperature for the minority populations is smaller, resulting in higher $T/T_F$ ratios. This leads to larger fluctuations in minority particles, increasing overall averaged local fluctuations.
Meanwhile, a larger $N$ suppresses fluctuation due to the interaction effect, which is consistent with our analysis of balanced systems. We obtain an analytical formula describing the feature,
\begin{align}
    \eta=&\frac{-3\pi}{4a_sk_F^\text{hom}\sum_\sigma\xi_\sigma}{\frac{T}{T_F^\text{hom}}}\dfrac{(\sum_\sigma g_\sigma)^2}{\sum_\sigma g_\sigma(1-\sum_{\sigma'\neq \sigma} g_{\sigma'})},
    \label{imbalanced_fluctuation}
\end{align}
where 
\begin{equation}
g_\sigma=\frac{a_s k_F^\text{hom}T^{\frac{1}{2}}}{\sqrt{\pi}(T_F^\text{hom})^{\frac{1}{2}}}\mathrm{Li}_{\frac{1}{2}}\left[\mathrm{Li}_{\frac{3}{2}}^{-1}\left(-\frac{4\xi_\sigma(T_F^\text{hom})^{\frac{3}{2}}}{3\sqrt{\pi}T^\frac{3}{2}}\right)\right]
\label{gsigma}
\end{equation}
with ${}^{-1}$ denotes the inverse function.
When $\xi_\sigma=1$ for all possible $\sigma$, Eq.~(\ref{imbalanced_fluctuation}) reduces to Eq.~(\ref{balanced_fluctuation}).
By comparing with experimental results, especially in Fig.~\ref{Imbalanced}(b) where the interaction effect is pronounced, Eq.~(\ref{imbalanced_fluctuation}) under local-density approximation with factor $\alpha$ is shown to explain the experimental data much better than calculations ignoring interaction effects.
The minor discrepancy observed near the $\operatorname{SU}(1)$ limit in Fig.~\ref{Imbalanced}(a), $\xi_\sigma -1\lessapprox 0.1$, may be attributed to two factors: unstable experimental control of the population ratio and the reduced accuracy of the Hartree-Fock approximation in this regime. This ``polaron physics" regime~\cite{massignan2014polarons} may be better described with more diagrams included.

After comparing with experimental data, we further emphasize the importance of Eq.~(\ref{imbalanced_fluctuation}) by presenting theoretical results in a homogeneous system. 
Figure~\ref{Imbalanced}(c) show the change of $\eta$  as the $\operatorname{SU}(1)$ gas is smoothly tuned to the $\operatorname{SU}(6)$ limit by changing $\xi_\sigma$ for $\sigma=2,\cdots,6$ from $0$ to $1$ \textit{one by one}. Solid and dashed lines in the upper panel show the fluctuation of interacting systems $\eta^\mathrm{int}$ and non-interacting systems $\eta^\mathrm{nonint}$, respectively. We observe that the difference increases as the total spin increases. Based on Eq.~(\ref{balanced_fluctuation}), the $(N-1) a_s$ interaction enhancement in spin-balanced $\operatorname{SU}(N)$ gases is also applicable to the inverse fluctuation, i.e., $1/\eta^\mathrm{int}-1/\eta^\mathrm{nonint}\propto (N-1)a_sk_F^\mathrm{hom}T_F^\mathrm{hom}/T$. The dashed lines (dotted lines in the lower panel) show our naive generalization $[(\sum_\sigma \xi_\sigma)-1)a_sk_F^\mathrm{hom}] T_F^\mathrm{hom}/T$ to the spin-imbalanced systems. In comparison, the solid lines show the inverse change of fluctuation based on Eq.~(\ref{imbalanced_fluctuation}). They agree well at high temperatures and differ slightly at low temperatures.

Changing the scheme to create the spin imbalance brings us even more interesting results. In Fig~\ref{Imbalanced}(d), we tune all $\xi_{\sigma>2}$ \textit{simultaneously} from $0$ to $1$ and show the same quantities. In the upper panel, we observe a clear increase in the difference of fluctuations of the interacting (solid lines) and non-interacting (dashed lines) systems at high temperatures ($T/T_F^\mathrm{hom}=2$ and $T/T_F^\mathrm{hom}=1$). However, in deep degeneracy regimes ($T/T_F^\mathrm{hom}=0.2$), the difference saturates at $\sum_\sigma \xi_\sigma\approx2$, where each minor species approximately has only $10\%$ of the major population. Similarly, in the lower panel, comparing to the change of inverse fluctuation in the previous case, we obverse a much prominent difference between the results based on Eq.~(\ref{imbalanced_fluctuation}) and the naive generalization, especially when the minor-major spin ratio is around $10\%$. This generalized $\operatorname{SU}(N)$ interaction enhancement suggests that compared to the interaction energy, the fluctuation is much more sensitive to the scattering length at sufficiently low temperatures, especially in highly imbalanced population setups. 

\begin{figure}[htb!]
	\centering
	\includegraphics[width=8.6cm]{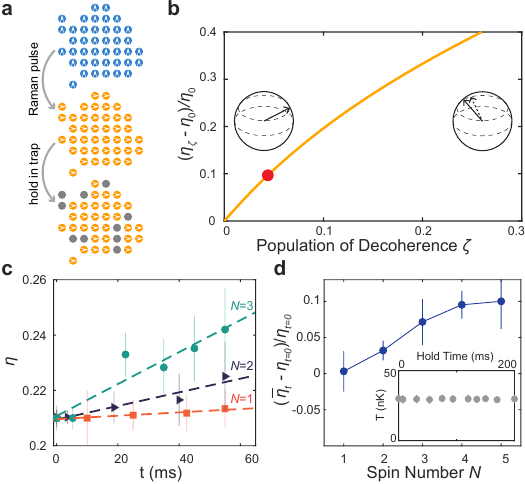}
	\caption{
    \textbf{Density fluctuation during spin decoherence.} 
    (a) A schematic showing the Raman pulse changes the gas in a single spin state to a coherent spin  state, and decoherence happens in the system.
    (b) Relative difference in variance per atom $\eta$ between the post-decoherence state $\eta_\zeta$ and initial state $\eta_0$ for quenching to $N=5$ at temperature $T=0.2T_F$ [based on Eq.~(\ref{imbalanced_fluctuation})]. The red circle denotes the point when the fluctuation increases by $10\%$. 
    (c) Experimental measurement of $\eta$ for spin-polarized gas quench to spin $N$ state ($N$ from 1 to 3). Dashed lines are linear fits.
    (d) Relative difference in variance per atom between the post-decoherence and initial averaged over a period from $40$ms to $100$ms. Inset: Circles show the fitted temperatures as a function of the holding time.
    }
	\label{quench_dynamics}
\end{figure}

\textit{Application of Fluctuation Measurement: Determining Decoherence.---}Understanding how an imbalanced population can affect density fluctuation allows us to reveal the decoherence dynamics in $\mathrm{SU}(N)$ Fermi gases~\cite{fletcher2017two,cetina2016ultrafast,bardon2014transverse,gupta2003radiofrequency}.
We first prepare a spin-polarized $\operatorname{SU}(1)$ Fermi gas in equilibrium, then apply a Raman pulse to suddenly rotate the spin to an equal superposition of $N$ spin states.
Initially, one expects each spin to be rotated to the same superposition state [Fig.~\ref{quench_dynamics}(a)], and the fluctuation should be kept the same as the $\operatorname{SU}(1)$ case.
In the long time limit, the fluctuation of gas approaches that of a $\operatorname{SU}(N)$ gas due to the spin decoherence [we observe that the heating effect induced by Raman pulse is negligible, see inset of Fig.~\ref{quench_dynamics}(d)].
In between, we interpolate the population ratios from $\operatorname{SU}(1)$ to $\operatorname{SU}(N)$, i.e., $(\xi_1,\xi_2,\cdots,\xi_N)=(1-\frac{N-1}{N}\zeta,\frac{\zeta}{N},\cdots,\frac{\zeta}{N})$, where $\zeta$ denotes the ratio of population of decoherence.
That is, we treat systems after decoherence as imbalanced $\operatorname{SU}(N)$ Fermi gases discussed previously and apply Eq.~(\ref{imbalanced_fluctuation}) with local-density approximation to calculate the fluctuations. 
Fig.~\ref{quench_dynamics}(b) shows our results for $N=5$. Indeed, we verify that as $\zeta$ increase, $\eta$ becomes larger at $T/T_F\sim 0.2$, where $T_F$ is calculated based on $N_0^{(1)}$ at $t=0$. 

Moreover, only a small proportion of decoherence can lead to relatively severe fluctuation change.  Experimentally, we measure the density fluctuation for different holding times after spin rotation. Squares, triangles, and circles with error bars in Fig.~\ref{quench_dynamics}(c) show the fluctuation for spin number $N=1,2,3$, respectively, as a function of holding time. We observe that the fluctuation increases faster when $N$ is larger. Circles in Fig.~\ref{quench_dynamics}(d) show the relative change of fluctuations averaged between $40~\text{ms}$ and $100~\text{ms}$ holding time as a function of spin number. We find a $10\%$ increase of $\eta$ after the Raman pulse for $N=5$. This $10\%$ change in variance corresponds to around $5\%$ atoms undergoing decoherence (the red circle in Fig.~\ref{quench_dynamics}(b)). Our estimation is qualitatively confirmed by double-pulse Ramsey 
spectroscopy, showing that most atoms in superposition states maintain coherence for up to $100\text{ms}$ (See End Note).

\textit{Conclusion.---} To summarize, we have examined density fluctuations in multi-component Fermi gases with $\mathrm{SU}(N)$ symmetric interactions. We have derived analytical formulas for calculating density fluctuations in general Fermi gases with $\mathrm{SU}(N)$ symmetric interactions and spin-imbalanced configurations. Our experimental measurement of density fluctuation with ${}^{173}\mathrm{Yb}$ atoms shows excellent agreement with theoretical predictions across a wide temperature range. As an application, we have demonstrated how density fluctuation measurements can be used to monitor decoherent populations. This establishes crucial fluctuation-based thermometry for precise measurements in multi-component Fermi gases with $\mathrm{SU}(N)$ symmetric interactions and opens new possibilities for exploring physics in imbalanced multi-component systems, such as a multi-component polaron problem. Furthermore, by combining orbital Feshbach resonance with oscillating magnetic fields~\cite{zhang2015orbital,cappellini2019coherent}, we could potentially achieve strong interactions between one spin component and several others. This approach could further enhance the interaction effects and help drive the gas into the superfluid regime.

\begin{acknowledgments}
\paragraph*{\bf Acknowledgement}
This work is supported by C4050-23G. GBJ also acknowledges support from the RGC through 16302420, 16302821, 16306321, 16306922, 16302123, C6009-20G, N-HKUST636-22, and RFS2122-6S04. YY acknowledges support from the RGC through 24308323, NSFC through 12204395, and the Space Application System of China Manned Space Program. CH acknowledges support from the RGC for RGC Postdoctoral fellowship.
\end{acknowledgments}

\clearpage
\appendix
\begin{widetext}

\section{Supplemntary Information}
\label{endnote}

\section{Atom Sample Preparation and Calibration}

\begin{figure}[b]
	\centering
	\includegraphics[width=0.44\textwidth]{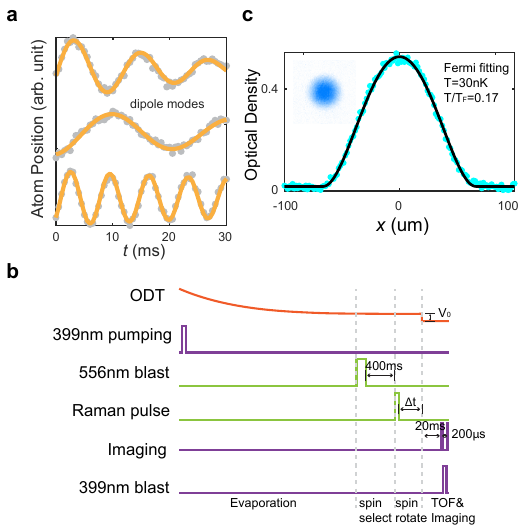}
	\caption{\textbf{Experimental sequence.} (a) ODT trap frequency calibration $\{(\omega_1,\omega_2,\omega_3)\} = \{84,46,143\}\times2\pi$ Hz. (b) The temperature of the atomic sample is controlled by varying final ODT depth $V_0$. Different spin configurations are prepared using $399~\text{nm}$ pumping and $556~\text{nm}$ blast pulse; atoms are held in ODT for $400~\text{ms}$ to reach thermal equilibrium before measurement. A Raman pulse is used to flip spin in quench dynamics measurement. Short inter-frame time absorption imaging is applied to measure atom distribution after $20~\text{ms}$ time-of-flight. The inter-frame time is $200~\mu\text{s}$, and a $399~\text{nm}$ blast pulse is applied to clean all atoms between two frames. (c) The atom cloud profile is fitted with the Fermi-Dirac distribution function to determine the temperature.}
	\label{experimental_sequence}
\end{figure}

\begin{figure}[hbt]
	\centering
	\includegraphics[width=0.44\textwidth]{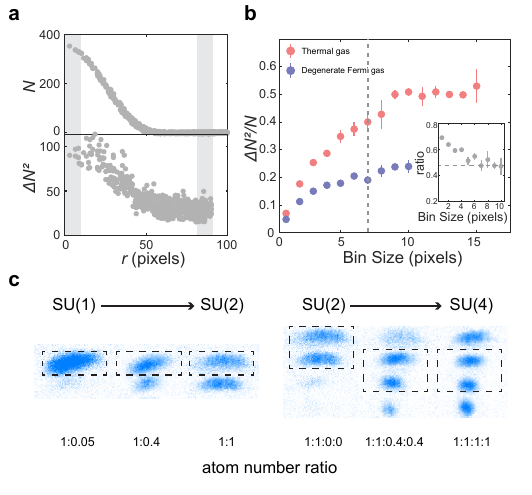}
	\caption{\textbf{Imaging system calibration and Optical Stern-Gerlach measurements} (a) The Upper panel shows the average atom number in one binning area, and the $x$ axis is the distance away from the center of the atom cloud. The lower panel is the total variance. At the area far from the center, where there are almost no atoms, total variance is a constant value contributed by photon shot noise, read-out noise, and dark noise. The shaded region on the left indicates density fluctuation at the center of the atom cloud; the shaded region on the right acts as a background. (b) Variance per atom when taking different bin sizes with the same datasets. When the bin size is large enough, variances tend to be constant. The inset plot shows the variance ratio between degenerate Fermi gas and thermal gas. (c) The atom number in each spin state is examined through Optical Stern-Gerlach measurement. We can prepare arbitrary spin configurations by tuning the spin pumping and blast beam power. The left part shows an imbalanced two-spin Fermi gas with a tunable atom number ratio. The right part shows a four-spin Fermi gas with two major and two minor spin components.}
	\label{binning_size}
\end{figure}

\begin{figure}[hbt]
	\centering
	\includegraphics[width=0.44\textwidth]{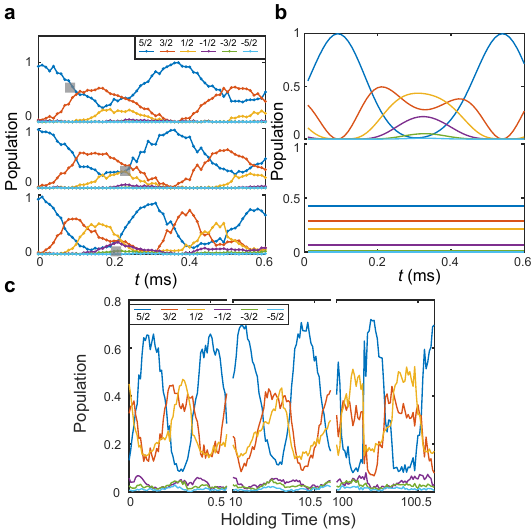}
	\caption{\textbf{Rabi oscillation during quench dynamics.} (a) Rabi oscillation for quench pulse from spin-polarized Fermi gases to near equally distributed 2-,3-,4-spin Fermi gases. We truncate Rabi oscillation at a gray square. Higher Raman pulse power is required for transitions to larger-spin-number states. (b) Theoretical predictions of atom population after $\pi/2$ pulse for a coherent gas (upper panel) and decoherent gas (lower panel) start from a $\operatorname{SU}(2)$ Fermi gas.  (c) Spin population measurement after two $\pi/2$ pulse for different inter-pulse holding time. Rabi frequency between $m_F = 3/2$ and $m_F = 5/2$ is around 5~kHz.}
	\label{rabi}
\end{figure}

We produce degenerate Fermi gas through evaporation cooling in ODT [Fig.~\ref{experimental_sequence}(b)]. The temperature of the atom sample is controlled by tuning the final ODT depth $V_0$. Different spin configuration samples are prepared by combining spin pumping and blast. Pumping beam wavelength is near $399~\text{nm}$ red detuned 400MHz from $^1$S$_0$ - $^1$P$_1$ transition with $\sigma+$ polarization. The spin blast beam is $556~\text{nm}$, resonance frequency of narrow line $^1$S$_0$ - $^3$P$_1$ transition with both $\sigma+$ and $\sigma-$ polarization. After preparing a specific spin configuration, we hold atoms in the trap for $400~\text{ms}$ to reach thermal equilibrium. We use a Raman beam, which is $1~\text{GHz}$ blue detuned from $^1$S$_0$ - $^3$P$_1$ transition with mix $\pi$ and horizontal linear polarization, to flip atom spin. We switch off ODT and let the atom cloud expand freely for $20~\text{ms}$ before absorption imaging. To reduce fringes in the absorption image, we use a short inter-frame time between the first and reference image within $200~\mu\text{s}$. We use a resonance $399~\text{nm}$ blast beam between two frames to blow the atom cloud away from the image area.

ODT trap geometry is calibrated by measuring dipole mode frequency as shown in Fig.~\ref{experimental_sequence}(a). We quench ODT trap depth to trigger the vertical motion of the atom cloud and use an optical-Stern-Gerlach beam to give atoms momentum kick in the horizontal direction. A typical value of our trap frequency is $\{(\omega_1,\omega_2,\omega_3)\} = \{84,46,143\}\times2\pi$ Hz. Fermi temperature is $T_F = \frac{\hbar}{k_B}\sqrt[3]{\omega_1\omega_2\omega_3\times6N_0^{(1)}}$. We extract temperature by fitting the atom cloud profile with Fermi-Dirac distribution as shown in Fig.~\ref{experimental_sequence}(c).

Noise in the absorption image from photon shot or camera readout noise will affect extraction atom number variance. To eliminate those effects, except for variance within ROI, we also extract the total variance for areas far away from the atom cloud as background, where almost no atom exists. We take the variance value after subtracting the background for each sample set. Fig.~\ref{binning_size}(a) shows a typical atom number and its variance versus distance from the atom cloud center with two gray areas marked as ROI and background area correspondingly. 

Because of the imperfection of the imaging system, binning size will affect the extracted variance value, as shown in Fig.~\ref{binning_size}(b). As we increase the binning size, the variance value increases and approaches a constant value. Considering the limited atom cloud size, we cannot use a binning size too large. 

By carefully tuning blast beam power for each spin state, we can prepare balanced and imbalanced spin configurations. The preparation result can be examined by counting atom number after OSG pulse as shown in Fig.~\ref{atom_variance} (c).

In the experiment, we quench atom spin using a Raman pulse. We rotate the spin of spin-polarized atoms to superposition states of $N$ spin. We can change the superposition states by tuning the power, the power ratio between $\pi$, and linear polarization and pulse time of the Raman beam. The coefficient of each spin component is nearly the same (within $10\%$ difference) as shown in Fig.~\ref{rabi}(a). We use a Ramsey spectroscopy to test the coherence of the atom sample. For a two-spin system, the first $\pi/2$ pulse will flip the spin from $|\uparrow\rangle$ to $(|\uparrow\rangle+|\downarrow\rangle)/\sqrt{2}$, after a waiting time $\delta t$, the second $\pi/2$ will rotate to a $\delta t$ dependent superposition states of spin up and spin down if there is no decoherence. For spin $N=6$ system, the first $\pi/2$ pulse will rotate spin from $|m_F = 5/2\rangle$ to $|m_F = 5/2\rangle+|m_F = 3/2\rangle$ usually with a small fraction of $|m_F = 1/2\rangle$ as plotted in Fig.~\ref{rabi}. The second pulse will rotate spin to the superposition of all six spin states depending on inter-pulse waiting time $\delta t$. If the system keeps fully coherent for a certain waiting time, the spin can return to $|m_F = 5/2\rangle$. However, if there is decoherence, the contrast of spin oscillation will drop as shown in Fig.~\ref{rabi}(b). If the system becomes fully decoherent, the final states after the second $\pi/2$ pulse will be time-independent. According to our spectroscopy up to $100~\mathrm{ms}$ hold time, the spin population oscillates as expected in a coherent sample (Fig.~\ref{rabi}(b)) with large contrast, which suggests the system remains coherent within $100~\mathrm{ms}$ (Fig.~\ref{rabi}(c)).

\section{Formalism of Calculating Density Fluctuation of Homogeneous Systems}

This section fully derives Eq.~(\ref{imbalanced_fluctuation}) from scratch in the main text. We work in a grand canonical ensemble, and the Hamiltonian of a general imbalanced multi-component Fermi gas with $\mathrm{SU}(N)$ symmetric interaction is given by
\begin{equation}
    K=\sum_{\sigma}\sum_{\mathbf{k}}\left(\frac{\hbar^2k^2}{2M}-\mu_\sigma\right)c_{\sigma,\mathbf{k}}^\dagger c_{\sigma,\mathbf{k}}+\frac{U}{2V}\sum_{\sigma}\sum_{\sigma'\neq\sigma}\sum_{\mathbf{P},\mathbf{q},\mathbf{q}'}c_{\sigma,\mathbf{P}-\frac{\mathbf{k}'}{2}}^\dagger c_{\sigma',\mathbf{P}+\frac{\mathbf{k}'}{2}}^\dagger c_{\sigma',\mathbf{P}+\frac{\mathbf{k}}{2}} c_{\sigma,\mathbf{P}-\frac{\mathbf{k}}{2}},
\end{equation}
where $\sigma=1,2,\cdots, N$ denotes the internal degree of freedom, $M$ is the mass of particles, and $U={4\pi \hbar^2 a_s}/{M}$ is the interaction strength expressed with $s$-wave scattering length $a_s$. 

\subsection{Hartree-Fock Approximation}
\label{appendix:B1}

In most experiments, including this work, the interaction is weak; thus, $a_s k_F$ is generally smaller than $1$, where $\hbar k_F$ denotes Fermi momentum. However, the unique feature of SU($N$) Fermi gas (let us assume the balanced case for the discussion here, $\mu_\sigma=\mu$) is, although the bare interaction is weak, the mean-field energy $V_\mathrm{MF}$ felt by each particle can be hefty since $V_\mathrm{MF}=(N-1)a_sk_F$ is enhanced by a factor of $N-1$~\cite{yip2014theory, cheng2017mathrm}. Consequently, when the interaction effect is investigated by diagrammatic expansion, it is necessary to include high-order contributions. Here, we show that the Hartree-Fock (HF) approximation is suitable for SU($N$) Fermi gas, which becomes exact in the large $N$ limit. 

\begin{figure}[h]
    \centering
    \includegraphics[width=0.49\textwidth]{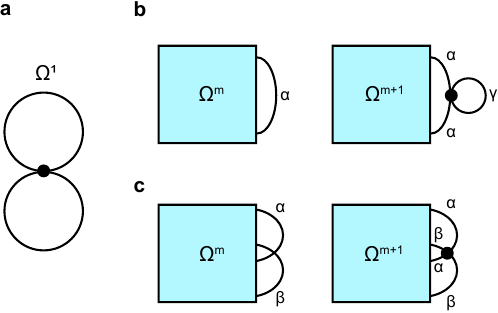}
    \caption{(a) First order correction due to interaction of the grand potential $\Omega$. Solid lines and the dot signify Green functions and interaction. (b) Generating the next order of the Feynman diagram by adding a new particle line and breaking the original one particle line into two. (c) Generating the next order of the Feynman diagram by breaking the original two particle lines into four.}
    \label{SUNenhancement}
\end{figure}

For $m$-order diagrams of the grand potential, they are enhanced by a factor $N(N-1)^{m}$ at most, i.e., except the particle line got lastly integrated has $N$ choices, other particle lines all have $(N-1)$ choices. However, the enhancement will generally be $N(N-1)^{m'}$ with $m'\leq m$ due to the topology of specific diagrams. For example, $m$-order ($m\geq2$) pair fluctuation is only enhanced by $N(N-1)$ since there are only two individual particles engaged in (consecutive) interactions. We can use the induction method to prove that diagrams enhanced by $N(N-1)^{m}$ are all HF types in each order. As shown by Fig.~\ref{SUNenhancement}(a), the first-order diagram of grand potential, according to the definition, belongs to the HF approximation. Let us assume the statement is correct for the $m$-order diagram. When we draw a $m+1$-order diagram based on a $m$-order diagram, we can either add a closed loop to a particular particle line [Fig.~\ref{SUNenhancement}(b)] or cross two particle lines to form a node [Fig.~\ref{SUNenhancement}(c)]. For the former operation, one has an additional $N-1$ internal degree freedom $\gamma$ to choose for the new loop. Thus, the new diagram will be enhanced by $N(N-1)^{m+1}$, which is indeed of HF type. In the latter case, the internal degree of freedom of the new particle line has already been determined by $\alpha$ and $\beta$. Thus, no additional enhanced factor is added. Such modifications are not on a single particle line, so the consequent diagram is not of HF type. Therefore, in the limit $N\rightarrow\infty$, HF terms in the diagrammatic expansion will dominate, and HF approximation nearly produces the exact result. 

Following the above discussion, we write the grand potential $\Omega$ of the system to be~\cite{fetter2012quantum}
\begin{equation}
    \Omega-\Omega_0=\frac{V}{2}\sum_{\sigma}\int_0^1\frac{d\lambda}{\lambda}\hbar\Sigma_\sigma^*(\lambda)n_{\sigma}(\lambda).
    \label{grandpotential}
\end{equation}
$\Omega_0$ is the grand potential for a non-interacting system
\begin{equation}
    \Omega_0=k_B T V \sum_{\sigma}\frac{\mathrm{Li}_{5/2}\left(-z_{\sigma}\right)}{\lambda_T^3},
\end{equation}
and $\Sigma_\sigma^*(\lambda)$ is the proper self-energy with interaction strength multiplied by $\lambda$
\begin{equation}
    \hbar \Sigma^*_\sigma(\lambda)=(2\pi)^{-3}\lambda U\sum_{\sigma'\neq\sigma}\int d^3k n_{\sigma',k},
    \label{HFselfenergy}
\end{equation}
where the self-consistent condition is~\cite{fetter2012quantum}
\begin{equation}
    n_{\sigma,k}=\dfrac{1}{e^{\beta\epsilon_{\sigma,k}}z_{\sigma}^{-1}+1},~~~\epsilon_{\sigma,k}=\frac{\hbar^2k^2}{2M}+\hbar\Sigma_\sigma^*.
    \label{selfconsistentcondition}
\end{equation}
The fugacity $z_\sigma$ used in the above formulae is defined by $z_\sigma=\exp[\beta \mu_\sigma]$. Combining Eqs.~(\ref{HFselfenergy}) and (\ref{selfconsistentcondition}), the self-consistent equation for density for $\lambda=1$ is
\begin{equation}
    n_\sigma \lambda_T^3+\mathrm{Li}_{3/2}\left[-\exp\left(-2a_s\lambda_T^2\sum_{\sigma'\neq\sigma}n_{\sigma'}\right)z_\sigma\right]=0,
    \label{selfconsistentn}
\end{equation}
where $\lambda_T$ is thermal wavelength defined by $\lambda_T=\hbar\sqrt{2\pi\beta/M}$.

\subsection{Density Fluctuation}
\label{appendix:B2}

The starting point is the definition of density fluctuation,
\begin{equation}
    \Delta N^2_0=\langle N^2_0\rangle-\langle N_0\rangle^2=\sum_\sigma k_B T \frac{\partial N_0^{(\sigma)}}{\partial \mu_\sigma}+\sum_\sigma\sum_{\sigma'\neq\sigma}k_B T \frac{\partial N_0^{(\sigma)}}{\partial \mu_{\sigma'}}.
\end{equation}
Using density and fugacity to rewrite the formula,
\begin{equation}
    \Delta n^2=\sum_\sigma z_\sigma\frac{\partial n_{\sigma}}{\partial z_\sigma}+\sum_\sigma\sum_{\sigma'\neq\sigma}z_{\sigma'}\frac{\partial n_{\sigma}}{\partial z_{\sigma'}}.
    \label{deltansquare}
\end{equation}
For the convenience of the following discussion, we define
\begin{equation}
    n_\sigma=f_\sigma\left(\sum_{\sigma'\neq\sigma} {n}_{\sigma'}\right).
\end{equation}
Based on Eq.~(\ref{selfconsistentn}), $f_\sigma$ is
\begin{equation}
    f_\sigma(x)=-\frac{1}{\lambda_T^3}\mathrm{Li}_{3/2}\left[-\exp\left(2a_s\lambda_T^2x\right)z_\sigma\right].
\end{equation}
Then,
\begin{equation}
    \frac{\partial n_\sigma}{\partial z_{\sigma'}}=\\
    \begin{cases}
        \dfrac{\partial f_\sigma}{\partial z_\sigma}+\sum_{\sigma''\neq\sigma}\dfrac{\partial f_\sigma}{\partial n_{\sigma''}}\dfrac{\partial n_{\sigma''}}{\partial z_{\sigma}}~(\sigma'=\sigma)\\
        \sum_{\sigma''\neq\sigma}\dfrac{\partial f_\sigma}{\partial n_{\sigma''}}\dfrac{\partial n_{\sigma''}}{\partial z_{\sigma'}}~(\sigma'\neq\sigma)
    \end{cases}.
\end{equation}
An important observation here is
\begin{equation}
    \frac{\partial f_\sigma}{\partial n_{\sigma''}}=\left.\frac{\partial f_\sigma}{\partial x}\right|_{x=\sum_{\sigma'\neq\sigma} {n}_{\sigma'}}\frac{\partial ( \sum_{\sigma'\neq\sigma} {n}_{\sigma'})}{\partial {n}_{\sigma''}}=\left.\frac{\partial f_\sigma}{\partial x}\right|_{x=\sum_{\sigma'\neq\sigma} {n}_{\sigma'}}~(\sigma'\neq\sigma''),
\end{equation}
which has no dependence on $\sigma''$. Thus, for convenience, we use the following definition:
\begin{align}
     g_\sigma&=\dfrac{\partial f_\sigma}{\partial n_{\sigma''}}~(\sigma\neq\sigma''),\\
    h_\sigma&=\dfrac{\partial f_\sigma}{\partial z_\sigma}.
\end{align}
We can rewrite the above formula as follows:
\begin{equation}
        \frac{\partial n_\sigma}{\partial z_{\sigma'}}=\\
    \begin{cases}
        h_\sigma+g_\sigma \sum_{\sigma''\neq\sigma}\dfrac{\partial n_{\sigma''}}{\partial z_{\sigma}}~\text{for $\sigma'=\sigma$}\\
        g_\sigma \sum_{\sigma''\neq\sigma}\dfrac{\partial n_{\sigma''}}{\partial z_{\sigma'}}~\text{for $\sigma'\neq\sigma$}
    \end{cases}.
    \label{dnsigmadzsigmap}
\end{equation}
One should note in the second case, it is possible that $\sigma''=\sigma'$. Utilizing Eq.~(\ref{deltansquare}), we express the density fluctuation $\Delta n^2$ to be
\begin{equation}
    \Delta n^2=\sum_\sigma z_\sigma h_\sigma +\sum_\sigma \sum_{\sigma'\neq\sigma} z_\sigma (g_\sigma+1)\dfrac{\partial n_{\sigma}}{\partial z_{\sigma'}}
    \label{imbalanced_fluctuation_SM}
\end{equation}
The problem is turned to obtained the expression for $\frac{\partial n_{\sigma'}}{\partial z_\sigma}(\sigma'\neq\sigma)$. 
From Eq.~(\ref{dnsigmadzsigmap}), (we default regard $\sigma'\neq\sigma$ in below derivation), we obtain
\begin{equation}
    (1-g_\sigma g_{\sigma'})\frac{\partial n_{\sigma'}}{\partial z_\sigma}-(g_{\sigma'}+g_\sigma g_{\sigma'})\sum_{\sigma''\neq\sigma',\sigma''\neq\sigma}\frac{\partial n_{\sigma''}}{\partial z_{\sigma}}=h_\sigma g_{\sigma'}
\end{equation}
By swapping dummy indices $\sigma$ and $\sigma'$, then setting $\sigma'=N$, we obtain a linear equation
\begin{equation}
 \begin{bmatrix}
   a_1& b & \cdots & b \\
   b & a_2 & \cdots & b\\
   \vdots  & \vdots  & \ddots & \vdots  \\
   b & b & \cdots & a_{N-1} 
 \end{bmatrix}
 \begin{bmatrix}
     x_{1}\\
     x_{2}\\
     \vdots\\
     x_{N-1}
 \end{bmatrix}
 =
 \begin{bmatrix}
    c\\
    c\\
    \vdots\\
    c
 \end{bmatrix}
 \label{symmetriceqn}
\end{equation}
where $a_j=\frac{1-g_jg_N}{gj},~b=-\frac{1+g_N}{h_N},~x_j=\frac{\partial n_j}{z_N}$, and $c=1$.
The solution can be obtained by using Cramer's rule
\begin{equation}
    x_i=\det(A_i)/\det(A),
\end{equation}
where $A$ is the coefficient matrix in Eq.~(\ref{symmetriceqn}) and $A_i$ is the matrix with $i$-th column replaced by $(c,c,c,\cdots,c)^T$. It is straightforward to show
\begin{equation}
    \det(A_i)=c\prod_{j\neq i}(a_j-b).
\end{equation}
Thus
\begin{equation}
    x_i=\frac{c}{\det(A)}\prod_{j\neq i}(a_j-b).
    \label{imbalanced_diffeqn}
\end{equation}
Explicitly,
\begin{equation}
\dfrac{\partial n^{\sigma}}{\partial z_{N}}={h_N \prod_{\substack{\sigma'\neq \sigma\\\sigma'\neq N}}  \frac{1+g_{\sigma'}}{g_{\sigma'}}}/\left\{\prod_{{\sigma'}\neq N} \frac{1-g_{\sigma'} g_N}{g_{\sigma'}}-\bigg[\sum_{m=2}^{N-1}\times(m-1)({1+g_N})^m\sum_{S\in\{\{g_{\sigma'}|{\sigma'}\neq N\}_{m-2}\}}\prod_{g_{\tau}\in S}\left(\frac{1-g_\tau g_N}{g_\tau}\right)\bigg]\right\},
\end{equation}
where $\{\{g_i\}_{m}\}$ signifies all $m$-combinations of the set containing all $g_i$ (i.e. $\{g_1,g_2,\cdots,g_N\}$). For example, if $N=3$,
 \begin{equation}
     \begin{split}
         \{\{g_i\}_{1}\}&=\{\{g_1\},\{g_2\},\{g_3\}\},\\
         \{\{g_i\}_{2}\}&=\{\{g_1,g_2\},\{g_2,g_3\},\{g_3,g_1\}\},\\
         \{\{g_i\}_{3}\}&=\{\{g_1,g_2,g_3\}\}.
     \end{split}
 \end{equation}
Similarly, we can solve all the $\frac{\partial n_{\sigma}}{\partial z_{\sigma'}}$.
Substituting back to Eq.~(\ref{imbalanced_fluctuation_SM}), we know the fluctuation for homogeneous systems. After tedious algebraic simplification, we find
\begin{equation}
    \eta=-\frac{3\pi}{4a_sk_F^\mathrm{hom}\sum_\sigma\xi_\sigma}{\frac{T}{T_F^\mathrm{hom}}}\dfrac{\sum_{m=1}^{N}m\sum_{S\in\{\{g_{\sigma'}\}_{m}\}}\prod_{g_\tau\in S}g_\tau}{1-\sum_{m=2}^{N}(m-1)\sum_{S\in\{\{g_{\sigma'}\}_{m}\}}\prod_{g_\tau\in S}g_\tau},
\label{imbalancedfluctuation}
\end{equation}
where $T_F^\mathrm{hom}, k_F^\mathrm{hom}$ are the Fermi temperature/momentum for the \textit{first} species. For example, explicit forms for $N=2$ and $N=3$ are
\begin{equation}
    \eta(N=2)=\frac{-3\pi}{4a_sk_F^\mathrm{hom}(\xi_1+\xi_2)}{\frac{T}{T_F^\mathrm{hom}}}\frac{g_1+g_2+2g_1g_2}{1-g_1g_2}
    \label{imbalancedfluctuation_SU2}
\end{equation}
and
\begin{equation}
    \eta(N=3)=\frac{-3\pi}{4a_sk_F^\mathrm{hom}(\xi_1+\xi_2+\xi_3)}{\frac{T}{T_F^\mathrm{hom}}}\frac{g_1+g_2+g_3+2g_1g_2+2g_1g_3+2g_2g_3+3g_1g_2g_3}{1-g_1g_2-g_1g_3-g_2g_3-2g_1g_2g_3},
    \label{imbalancedfluctuation_SU3}
\end{equation}
where $g_\sigma$ is given in Eq.~(\ref{gsigma}) in the main text.

In spin-balanced systems, the interaction only appears up to the first order in the denominator of Eq.~(\ref{balanced_fluctuation}). This occurs because the unperturbed eigenfunctions in these systems are also the self-consistent eigenfunctions~\cite{fetter2012quantum}. However, this argument does not hold for spin-imbalanced systems, resulting in higher-order contributions in Eq.~(\ref{imbalancedfluctuation}) [as demonstrated explicitly in Eqs.~(\ref{imbalancedfluctuation_SU2}) and (\ref{imbalancedfluctuation_SU3})]. This raises the question: to what extent do these higher-order terms impact the results? To compare with spin-balanced systems, we calculate the lowest-order Pad\'e approximant of Eq.~(\ref{imbalancedfluctuation}), yielding Eq.~(\ref{imbalanced_fluctuation}) in the main text. Figure~\ref{relative_error} illustrates the relative value difference between these two expressions for the scenario depicted in the left panel of Fig.~\ref{Imbalanced}(b) in the main text. The differences between the expressions are minimal, with the most considerable discrepancy not exceeding $0.1\%$. Consequently, we can confidently use Eq.~(\ref{imbalanced_fluctuation}) in our calculations, significantly simplifying the combinatorics while maintaining accuracy.

\begin{figure}[h]
    \centering
    \includegraphics[width=0.45\textwidth]{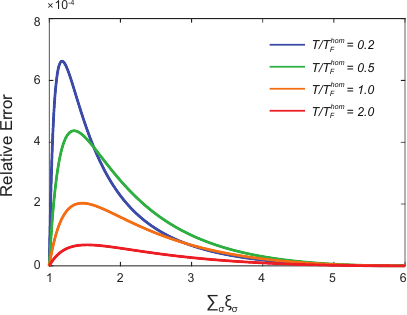}
    \caption{The relative error between the exact result Eq.~(\ref{imbalancedfluctuation}) and Pad\'e approximant Eq.~(\ref{imbalanced_fluctuation}) in the main text. Here, we set $a_s k_F^\mathrm{hom}=0.1$, and vary the population ratio by the scheme of the left panel of Fig.~\ref{Imbalanced}(b) in the main text.}
    \label{relative_error}
\end{figure}

\section{Line-of-Sight Integration}
\label{appendix:C}

For applying Eq.~(\ref{imbalancedfluctuation}) to experimental harmonically trapped systems, we perform the local density approximation (LDA) and integrate the result along the line-of-sight direction of imaging~\cite{sanner2010suppression}. Under LDA and according to Eqs.~(\ref{HFselfenergy}) and (\ref{selfconsistentcondition}), the phase space density for species $\sigma$ is given by
\begin{equation}
\begin{split}
    f_\sigma(\mathbf{x},\mathbf{k})={1}/\left\{\exp\left[\frac{\beta\hbar^2k^2}{2M}\right]\exp\left[\frac{\beta M\sum_i \omega_i^2 x_i^2}{2}\right]\exp\left[\frac{4\beta\pi\hbar^2a_s\sum_{\sigma'\neq\sigma}n_{\sigma'}(x_i)}{M}\right]z^{-1}+1\right\},
    \label{localPSD}
\end{split}
\end{equation}
where
\begin{equation}
    n_{\sigma}(\mathbf{x})=\int \frac{d^3k}{(2\pi)^3}f_\sigma(\mathbf{x},\mathbf{k})
\end{equation}
is the local density. Integrating $\mathbf{k}$ in Eq.~(\ref{localPSD}) gives self-consistent equations of $n_{\sigma}(\mathbf{x})$. For convenience, we express all quantities in their dimensionless form following:
\begin{align*}
    &(\omega_1,\omega_2,\omega_3)=\omega(\eta_1=1,\eta_2,\eta_3),\\
    &x_i\rightarrow\bar{x}_i R_F,~~~R_F=(48\eta_2\eta_3 N_0^{(1)})^{1/6}\sqrt{\frac{\hbar}{M\omega}},\\
    &k_B T\rightarrow \Tilde{T}E_F,~~~E_F=\hbar\omega(6\eta_2\eta_3 N_0^{(1)})^{1/3},\\
    &k\rightarrow \Tilde{k}k_F,~~~k_F=\dfrac{\sqrt{2ME_F}}{\hbar}=(48\eta_2\eta_3 N_0^{(1)})^{1/6}\sqrt{\frac{M\omega}{\hbar}},\\
    &X_i=\eta_i \Bar{x}_i,~~~n_\sigma(\mathbf{x})\rightarrow \bar{n}_\sigma(\mathbf{X})/R_F^3
\end{align*}
We obtain
\begin{equation}
    \frac{\Bar{n}_\sigma(\mathbf{X})}{N_0^{(1)}}=-\left(\prod_i \eta_i\right)\frac{6\Tilde{T}^{3/2}}{\pi^{3/2}}\mathrm{Li}_{3/2}\left(-\exp\left[-\sum_{\sigma'\neq\sigma}\frac{\pi\Tilde{a}_s\Bar{n}_{\sigma'}(\mathbf{X})/N_0^{(1)}}{6\Tilde{T}\prod_i \eta_i}\right]\exp\left[-\frac{X^2}{\Tilde{T}}\right]z_\sigma \right).
    \label{imbalanced_n}
\end{equation}
Differentiating Eq.~(\ref{imbalanced_n}) against $z_\sigma$ on both sides, we get a linear equation with structure Eq.~(\ref{symmetriceqn}), where
\begin{align*}
    a_j&=\frac{\pi^{3/2}}{12\sqrt{\Tilde{T}}\mathrm{Li}_{1/2}\left(\mathrm{Li}_{3/2}^{-1}\left[-\frac{\pi^{3/2}\Bar{n}_j(X)}{6 N_0^{(1)} \prod_i \eta_i \Tilde{T}^{3/2}}\right]\right)},\\
    b&=-\frac{\pi \Tilde{a}_s}{12},\\
    c&=X,\\
    x_j&=\frac{1}{N\prod_i \eta_i}\frac{d \Bar{n}_j(X)}{dX}.
\end{align*}
Solution Eq.~(\ref{imbalanced_diffeqn}) immediately gives $N$ first-order differential equations for convenient numerical work. Furthermore, we should determine $z_\sigma$ before solving the equations; equivalently, we need to input the boundary conditions $\Bar{\mathbf{n}}(0)=(\Bar{{n}}_1(0),\Bar{{n}}_2(0),\cdots,\Bar{{n}}_N(0))$. The boundary condition is implicitly given by
\begin{equation}
    \int d^3 X \frac{\Bar{n}_\sigma(X)}{N_0^{(1)} \prod_i\eta_i}=\xi_\sigma.
\end{equation}
To find the explicit boundary condition, we adopt the generalization of the Newton-Raphson iteration method in higher dimensions. In each step of the iteration, the update is
\begin{equation}
\begin{split}
    &\Bar{\mathbf{n}}_{\mathrm{new}}(0)=\Bar{\mathbf{n}}(0)+\Delta\mathbf{\Bar{n}}(0),\\
    &J_\mathcal{F}\left[\frac{\mathbf{\Bar{n}}(0)}{N_0^{(1)}\prod_i\eta_i}\right]\frac{\Delta\mathbf{\Bar{n}}(0)}{N_0^{(1)}\prod_i\eta_i}=-\mathcal{F}\left[\frac{\mathbf{\Bar{n}}(0)}{N_0^{(1)}\prod_i\eta_i}\right],
\end{split}
\end{equation}
where 
\begin{equation}
    \mathcal{F}[\frac{\mathbf{\Bar{n}}(0)}{N_0^{(1)}\prod_i\eta_i}]=\int d^3X \frac{\mathbf{\Bar{n}}(\mathbf{X})[\mathbf{\Bar{n}}(0)]}{N_0^{(1)}\prod_i \eta_i}-\Vec{\xi}
\end{equation}
and $J_\mathcal{F}$ is the Jacobian matrix of $\mathcal{F}$. After solving $\mathbf{\Bar{n}}(\mathbf{X})$, we can obtain the total fluctuation by integrating fluctuation along the line of sight (in this work, the imaging is along $y=x/\sqrt{3}$. However, arbitrary imaging directions give same results since the factor will be canceled out later)
\begin{equation}
    \frac{\Delta N^2}{N}=\frac{\int d x_1 \Delta n^2(x_1,x_1/\sqrt{3},0)}{\int d x_1 n(x_1,x_1/\sqrt{3},0)},
\end{equation}
$\Delta n^2(\mathbf{x})$ can be obtained by LDA of Eq.~(\ref{imbalancedfluctuation}). Converting the expression of denominator and numerator to quantities in the dimensionless form, we have
\begin{equation}
    \int d x_1 n(x_1,\frac{x_1}{\sqrt{3}},0)=\frac{\sum_\sigma}{R_F^2 \sqrt{\eta_1^2+\eta_2^2/3}}\int dX  \Bar{n}_\sigma(X),
\end{equation}
and
\begin{equation}
\begin{split}
    &\int d x_1 \Delta n^2(x_1,\frac{x_1}{\sqrt{3}},0)=\frac{-1}{R_F^2 \sqrt{\eta_1^2+\eta_2^2/3}}\\
    &\times\int dX \dfrac{6\Tilde{T} N_0^{(1)} \prod_i \eta_i }{\pi \Tilde{a}_s}\\
    &\times\dfrac{\sum_{m=1}^{N}m\sum_{S\in\{\{G_\sigma\}_{m}\}}\prod_{G_\tau\in S}G_\tau(X)}{1-\sum_{m=2}^{N}(m-1)\sum_{S\in\{\{G_\sigma\}_{m}\}}\prod_{G_\tau\in S}G_\tau(X)},
\end{split}
\end{equation}
where
\begin{equation}
    G_\sigma(X)=\sqrt\frac{\Tilde{T}}{\pi}\Tilde{a}_s\mathrm{Li}_{1/2}\left[\mathrm{Li}_{3/2}^{-1}\left(-\frac{\pi^{3/2}\Bar{n}_\sigma(X)}{6\Tilde{T}^{3/2}N_0^{(1)}\prod_i\eta_i}\right)\right].
\end{equation}
Finally, we have
\begin{equation}
     \eta=\dfrac{\int dX \dfrac{6\Tilde{T}}{\pi \Tilde{a}_s}\dfrac{-\sum_{m=1}^{N}m\sum_{S\in\{\{G_\sigma\}_{m}\}}\prod_{G_\tau\in S}G_\tau(X)}{1-\sum_{m=2}^{N}(m-1)\sum_{S\in\{\{G_\sigma\}_{m}\}}\prod_{G_\tau\in S}G_\tau(X)}}{\int dX \sum_\sigma \frac{\Bar{n}_\sigma(X,0)}{N_0^{(1)}\prod_i \eta_i}}.
\end{equation}
\end{widetext}

%

\end{document}